# Gravitation as a pressure force: a scalar ether theory

**Mayeul Arminjon**
Laboratoire "Sols, Solides, Structures", Institut de Mécanique de Grenoble
B.P. 53, 38041 Grenoble Cedex 9, France

## 1. Introduction and summary

The concept of an ether means primarily that empty space does not really exist. We may believe this, for instance, because electromagnetic waves, that go accross intergalactical space, ought to wave in some medium. We may also believe this, because quantum phenomena, such as the Casimir effect, suggest that "vacuum" actually has physical properties. It has been established by Builder [10-11], Jánossy [20-21], Prokhovnik [33-34], and others, that the concept of the ether as an inertial frame which should be the carrier of the electromagnetic waves (the Lorentz-Poincaré ether), is fully compatible with Special Relativity (SR). In connection with this, Zhang [41] has recently reestablished, against contrary statements, that the one-way velocity of light cannot be consistently measured – in the absence of any faster information carrier. As emphasized by Duffy [17], the Builder-Prokhovnik reconstruction of standard SR from the Lorentz-Poincaré ether concept may be criticized on the ground that this construction makes undetectable the absolute reference frame and its velocity, which are the physical entities with which the construction starts. It would not be an appropriate answer to recall that, after all, this is the way in which Lorentz, Larmor and Poincaré themselves derived the major part of SR: indeed, this methodological oddness – which is not a logical fault, however – contributed to bring discredit on the ether concept for a long time. Another possible answer would be to insist that, beyond physical concepts, one may still introduce metaphysical ones.

The answer that will be suggested here is, instead, that SR does not rule the whole of physics, for it does not involve gravitation. If the presence of a gravitational field breaks the symmetry expressed by the Lorentz group, then the Lorentz-Poincaré construction of SR is justified as the examination of a particular physical situation, the general situation (with gravitation) being one in which the "absolute motion" should be detectable. Although it is admittedly a risk to build a preferred-frame theory of gravitation, it will be shown that the risk is smaller than one would think, if one would blindly follow the standard arguments. Moreover, there are well-known physical reasons that may justify to search for a radical alternative to general relativity: e.g. the singularity occuring during gravitational collapse, the problem of dark matter, and the questions about the influence of the gauge condition.

The theory summarized here starts from a tentative interpretation of gravity as a pressure force, which would give a new theory already in a *Newtonian space-time* ; this will be recalled in **Section 2**. However, to become compatible with SR, this interpretation of gravity calls for a new version of Einstein's equivalence principle, which consists in admitting that our physical standards of space and time are affected by gravitation in much the same way as they are affected by a uniform motion (**Section 3**). Yet in the "pseudo-Riemannian space-time" which is thus obtained, motion is governed by an *extension of Newton's second law*, instead of assuming Einstein's motion along geodesic lines of the space-time metric (**Section 4**). The extended Newton law implies an energy balance for a test particle moving in a variable gravitational field; this balance equation may be translated so as to apply to a dust, of course (dust is a continuum made of non-interacting particles). The field equation of the theory allows to rewrite the balance equation for a dust as a *true conservation equation for the total energy*, including the gravitational energy; this conservation equation is then assumed valid for any kind of material or non-gravitational field, characterized by its energy-momentum tensor (**Section 5**). Using a different method, one may obtain the 4-component equation governing the dynamics of a dust continuum in terms of the energy-momentum tensor; this equation also may be assumed to be valid for any continuum, it is the substitute for the classical equation $T^{\mu\nu}{}_{;\nu} = 0$ of general relativity and other metric theories. The new equation implies that *mass conservation is obtained as a limit behaviour for a weak and slowly varying gravitational field and/or at low pressure* (**Section 6**). The same method as in Sect. 6 may be used in the presence of a field of (non-gravitational) external force. When the external force is the Lorentz force, the 4-component dynamical equation thus derived gives the second group of the *gravitationally-modified Maxwell equations*, in the investigated theory. This new modification of Maxwell's equations is *consistent with the geometrical optics of the theory* as governed by the proposed extension of Newton's second law (**Section 7**). An essential question is *whether the theory agrees with experiment or not*, of course. The theory has the correct Newtonian limit; it predicts Schwarzschild's exterior metric of general relativity and geodesic motion in the static situation with spherical symmetry. Therefore, the question amounts principally to assessing the preferred-frame effects. To do this, one must develop a "post-Newtonian" approximation of this non-linear theory (**Section 8**). The first result is that *no preferred-frame effect occurs for photons at the (first) post-Newtonian approximation*. However, the absolute velocity of the solar system does influence the motion of mass points at this same approximation, thus it does have to be accounted for in celestial mechanics, according to the present theory. But it will be argued, contrary to well-known arguments, that *the existence of preferred-frame effects in celestial mechanics, comparable in magnitude with the "relativistic" effects, does not a priori invalidate a theory*.

## 2. Semi-heuristic considerations and a theory for Newtonian absolute space and time

Our starting point is that the concept of an ether should first be made compatible with *classical mechanics* [1]. This is not so easy: on the one hand, within classical mechanics, the only medium that could "fill the vacuum without us feeling it", would be a perfect fluid, which indeed would not brake any motion. But, on the other hand, we want that the ether defines an inertial frame, for otherwise we would have to postulate both an ether and an independent absolute space, and it seems that this would be too much of "absolute". (Moreover, the ether defines an inertial frame in the Lorentz-Poincaré interpretation of SR, although obviously this argument is less closely related to classical mechanics.) Hence, in the case of usual Newtonian mechanics, our perfect fluid should be perfectly rigid! The proposed answer is that *it is the average motion of the perfect fluid that defines the preferred inertial frame* [1-2] (the average should be taken at a very large scale, formally it is the asymptotic volume average [4]). It is worth to note that, even if one remains within classical mechanics, thus with our physical standards being assumed to measure Euclidean distance and absolute time, this preferred inertial frame does not actually need to be rigid. In other words, Newtonian mechanics may be extended to *deformable* inertial frames: the only restriction, imposed by the principle of inertia, is that the motion of the inertial frame with respect to a (any) rigid reference frame should be irrotational [1]. This generalization allows some simple cosmological considerations, in particular it gives a simple argument for an expanding universe [1]. An expanding preferred frame was already envisaged by Prokhovnik [33-34]. The motion (deformation) of the preferred inertial frame E, thus the average motion of the fluid ether, does not obey this "neo-classical" mechanics, since this average motion defines the frame of this mechanics. (It means that this mechanics does not say us whether E has a rigid motion, an expansion, or any other irrotational motion: we have to empirically determine this motion. Note that the "neo-classical" mechanics is the usual classical mechanics only in the case that the frame E has a rigid motion with respect to a (any) rigid reference frame.)

However, we may tentatively assume that the *microscopic* motion of the (micro-)ether *with respect to its mean rest frame* E or macro-ether, does obey mechanics, as well as the motion of material particles. The micro-ether being a perfect fluid, it exerts only surface forces due to its pressure, which may be reexpressed as volume forces depending on the gradient of the pressure of the micro-ether. Thus, any material, "elementary" particle, supposed to occupy a finite volume, is subjected to this Archimedes thrust, just like a ball in the sea (in this tentative reasoning, we deliberately "forget" the fact that quantum, not classical mechanics, is usually admitted to apply at this scale). In order that this pressure force be a universal force depending only on the inertial mass of the material particles, it is sufficient to assume that the mass density inside any particle (averaged over the volume of the particle) is the same for all particles, say $\rho_p$. Then a field $p_e$ of "ether pressure" would give the gravity acceleration $\mathbf{g} = -\,\mathrm{grad}\ p_e/\rho_p$ [1, 3]. (It turns out that this interpretation of a universal gravitation force was proposed by Euler in 1746! See Whittaker [39].) We then note that $\rho_p$ may still depend on $p_e$, but on $p_e$ only, and that this would seem miraculous unless the material particles themselves are made of the hypothetical fluid – as is also suggested by the unity of matter and fields, and by the observed transformations of particles into different ones. According to this notion of a "constitutive ether", material particles would be organized local flows in the ether, which seems a very promising heuristics as regards the realistic interpretation of these objects of microphysics – as shown in detail by Romani [35]. Hence, we equate $\rho_p$ to $\rho_e$, the local "density" of the micro-ether, which must depend on $p_e$ only, i.e. a barotropic fluid (this is necessarily the case for a perfectly continuous fluid, for which entropy and temperature do not make sense [35]). We thus assume that

$$\mathbf{g} = -\,\mathrm{grad}\ p_e/\rho_e. \tag{1}$$

It is worth noting that Eq. (1) implies a *decrease* of the ether pressure $p_e$ and the ether density $\rho_e = \rho_e(p_e)$ in the direction of the gravitational attraction. Because the gravitation varies only over macroscopic distances, $p_e$ must be the macroscopic pressure in the ether; more precisely, grad $p_e$ must be the average of grad $p'_e$ over a macroscopic volume, with $p'_e$ the "true" microscopic pressure of the ether [1]. This leaves the possibility that the other interactions, which vary over much shorter length scales, could be explained as combined effects of the field $p'_e$ on the matter particles, the latter being seen as organized flows. We see that gravitation can be envisaged as a *correction* defined by Eq. (1), even in the case where the "mass density inside a matter particle" would not be the same for all particles. In other words, at the scale of the elementary particles, the universality of the gravitation force might be a question of definition. Moreover, the sequel of the development of the theory (Sect. 3) leads to admit that the compressibility of the (micro-)ether is $K = 1/c^2$ with $c$ the velocity of light, thus an extremely low compressibility. In this situation, it is natural to expect that the mass density is indeed nearly the same for all particles.

Equation (1), thus obtained semi-heuristically, is yet assumed to be an exact equation of the theory, which is a substitute for the Newtonian equation $\mathbf{g} = \mathrm{grad}\ U$. In a phenomenological point of view, "a theory is defined by the set of its equations" (as Hertz said about Maxwell's electromagnetic theory). The next step is to state the equation relating the field of macroscopic ether pressure, $p_e$, to the mass density $\rho$ (the latter is, of course, the usual density, $\rho = \delta m/\delta V$, with $\delta m$ the sum of the masses of the material particles involved in some volume element $\delta V$ ). To do this, we assume, as did Romani [35], that Newtonian gravitation (NG) corresponds to the limit case of an incompressible fluid. Indeed, gravitation propagates instantaneously in NG, whereas, if gravitation is a pressure force, it should propagate with the "sound" velocity in the fluid ether,

$$c_e(\rho_e) \equiv \sqrt{dp_e / d\rho_e} , \qquad (2)$$

which is infinite only for an incompressible fluid. In order to recover Poisson's equation for the gravity acceleration **g** (Eq. (1)), the field equation must be in the incompressible case:

$$\Delta p_e = 4\pi G \rho \rho_e. \qquad (3)$$

In the static situation, the propagation speed plays no role; it is thus natural to assume that, in the static situation, this same equation still applies also in the case where the fluid indeed has a compressibility. However, this means already a new, non-linear theory of gravitation, implying e.g. a perihelion shift for the orbit of a test particle in a static field with spherical symmetry [1]. In assuming that the ether is conserved, that the wave motion of the ether with respect to its mean rest frame obeys Newton's second law, and by adapting standard arguments of acoustics, one then finds that the field equation of the general case is [1]

$$\Delta p_e - \frac{1}{c_e^2} \frac{\partial^2 p_e}{\partial t^2} = 4\pi G \rho \rho_e. \qquad (4)$$

Thus, the concept of a fluid ether leads quite naturally to the notion that gravitation should propagate in essentially the same way as the most usual waves, i.e. acoustic and electromagnetic waves. Equation (4) is not Galileo-invariant, hence this theory with Newtonian space-time is a preferred-frame theory. NG and Galilean invariance are recovered as the limit case of an incompressible fluid.

### 3. Principle of equivalence between the metric effects of uniform motion and gravitation

The foregoing theory has to be adapted so as to account for the "relativistic" effects, here seen as resulting from the "real" Lorentz contraction [10-11, 20-21, 33-34]. Thus, these effects are essentially absolute metric effects of uniform motion. Einstein's principle of equivalence between inertial effects and effects of a true gravitational field is an extrapolation of a ("weak") principle of equivalence, valid in Newtonian theory as a simple consequence of the identity between inertial mass and passive gravitational mass. As it has been often discussed in the literature (see e.g. Fock [18]), these equivalence principles are valid only in an infinitesimal domain, such that the gravitational field, as well as the field of "fictitious" inertial forces due to the motion of the considered frame with respect to the inertial frame, may be considered as uniform. Now, in a relativistic theory, the infinitesimal domain must be in space-time, so that the uniformity should be true both in space and in time. But, in an infinitesimal domain of space-time, any motion may be considered as uniform, and a uniform motion can have none other effect as the metric effects due to the FitzGerald-Lorentz contraction of the measuring rods and the Larmor-Lorentz-Einstein dilation of the clock periods. Hence, we are lead to the conclusion that, if there can be an equivalence principle, this should give the correspondence between the metric effects of a uniform motion and certain metric effects that should appear as a consequence of a (locally) uniform gravitational field.

Under our assumption that a gravitational field is due to a variation in the ether pressure, or equivalently in the ether density, we indeed can see an analogy between a uniform motion and a gravitational field: due to the Lorentz contraction, the "apparent" density of the ether is modified by the uniform motion. More precisely, let us assume that an observer following the uniform motion, at constant velocity **V** with respect to the macro-ether E, can use the "true" simultaneity (of the frame E), which he indeed can use if he knows his absolute velocity **V** – this, in turn, being *a priori* possible in a preferred-frame theory which will lead to definite effects of the absolute motion, at least in the presence of a gravitational field. Then, a given volume $dV^0$ of ether (as evaluated with the rods of E) has, for this observer, a greater volume $dV = dV^0/\beta_\mathbf{V}$, with $\beta_\mathbf{V} = (1 - V^2/c^2)^{1/2}$, because his measuring rod is contracted in the ratio $\beta_\mathbf{V}$ in the direction **V**. The "mass" or rather the amount of ether is unchanged, for the mass increase with velocity concerns only material particles. Hence, due to the absolute motion, the apparent ether density is lowered, $\rho_{e\,\mathbf{V}} = \rho_e . \beta_\mathbf{V}$. Now the metric effects of uniform motion depend only on $\beta_\mathbf{V}$, thus they depend only on the ratio between the "apparent" density which might be "evaluated" by the moving observer, and the "true" density that he might evaluate if he were not in a motion: $\beta_\mathbf{V} = \rho_{e\,\mathbf{V}}/\rho_e < 1$. Similarly, a gravitational field implies that the local ether density $\rho_e$ is lower (see after Eq. (1)) than the ether density $\rho_e^\infty$ in remote regions that are free from gravitation. We are thus naturally lead to postulate that [2-3]:

(A) *In a gravitational field, material objects are contracted, only in the direction of the field* **g** (with **g** = − grad $p_e/\rho_e$ ), *in the ratio*

$$\beta = \rho_e / \rho_e^\infty < 1 , \qquad (5)$$

*where $\rho_e^\infty$ is the ether density at a remote point where no gravity is present, and the clock periods are dilated in the same ratio.*

This statement is made for objects and clocks bound with E ; if this is not so, one has to combine the metric effects due to motion and gravitation. Due to the contraction of measuring rods in the direction **g**, the physical space metric **g** in the frame E becomes a Riemannian one. The contraction of the rods (hence the dilation of the measured distances) occurs with respect to a "background" Euclidean metric $\mathbf{g}^0$, which is assumed to be bound with the macro-ether E. The latter assumption means that the components $g^0_{ij}$ are constant in coordinates bound to the preferred frame – i.e., space-time coordinates $(x^\mu)_{\mu = 0, ..., 3}$, such that each point bound to E has constant space coordinates $(x^i)_{i = 1, 2, 3}$. [1] In the same way, the dilation of the clock periods implies a contraction of the local time interval $dt_\mathbf{x}$, measured with a fixed clock (bound to E) in a gravitational field, as compared with the corresponding interval of the "absolute time", $dT$, thus

$$dt_\mathbf{x}/dT = \beta. \qquad (6)$$

---

[1] Latin indices vary from 1 to 3 (spatial indices), Greek indices from 0 to 3.

This assumption (A) gives a specific form to the space-time metric $\gamma$ [2]. In particular, the slowing down of clocks, as expressed by Eq. (6), implicitly assumes that the absolute time $T$ is a globally synchronized time coordinate [23] in the frame E, i.e., the components $\gamma_{0i}$ are zero in coordinates $(x^\mu)$ that are bound to the preferred frame and such that $x^0 = cT$, which means that a simultaneity is defined in the frame E as a whole [3, 5]. The property "$\gamma_{0i} = 0$" holds true after any coordinate transformation of the form

$$x'^0 = \phi(x^0), \quad x'^i = \psi^i(x^1, x^2, x^3). \tag{7}$$

These transformations, that both leave the frame unchanged and keep true the global synchronization [5], build a group which is important in the theory [7-8]. However, in order that the time-dilation (6) be univoquely defined, it is necessary to restrict oneself to the subgroup constituted by the merely spatial transformations [7-8]; this justifies the term "absolute time" for $T$. Note that $T$ is the time which is measured by a clock bound to the preferred frame and far enough from massive bodies so that no gravitational field is felt.

The author recently learned that a somewhat similar equivalence principle between metric effects of uniform motion and gravitation had previously been assumed, in the special case of a static gravitational field, by Podlaha & Sjödin [30] (*cf.* also Podlaha [29] and Sjödin [36]). However, the assumption of these authors is different from assumption (A) hereabove: the ratio assumed by Podlaha & Sjödin is $\beta' = (\rho_0/\rho)^{1/2}$, where $\rho_0$ is the ether density in an unaffected region and plays thus the same role as our $\rho_e^\infty$, so that one might write $\beta = 1/\beta'^2$ if the "ether density" $\rho_e$ (or $\rho$ for Podlaha and Sjödin) were one and the same field in both approaches. Yet for Podlaha & Sjödin [30], the ether density *increases* in the direction of the gravitational attraction, this increase being approximately determined by the (opposite of the) Newtonian potential, $-U = -GM/r$ for the static field produced by a spherical body with mass $M$: their equation (20) amounts to $\rho/\rho_0 = (1 - U/c^2)^{-2}$, or $\beta' = 1 - U/c^2$. For a weak gravitational field, our assumption (A) gives $\beta \approx 1 - U/c^2$ [2], thus $\beta \approx \beta'$ for a weak field. However, our assumption is stated for any gravitational field, and even for a weak field there are correction terms [6] (see Eq. (64) here). We have given a heuristic justification for assumption (A) (see also Refs. 2-3). Another difference is the way in which the spatial contraction is assumed to occur, i.e., whether it is isotropic or not: Podlaha & Sjödin do not precise this point, but Eq. (8) in Podlaha [29] seems to imply that the contraction would be locally isotropic. According to our hypothese, the spatial contraction is anisotropic (just like the FitzGerald-Lorentz contraction), since it occurs in the direction of the **g** vector only.

As a result of the proposed equivalence principle, two different space metrics, **g** and **g**$^0$, as well as two different times (the local time $t_\mathbf{x}$ and the absolute time $T$), might be used by observers bound to the preferred frame. Combining the space and time evaluations into a space-time metric (which, in contrast, is valid for any observer in any reference frame), we may also say that we have two such metrics on the "space-time": an "abstract" one, $\gamma^0$, obtained by combining **g**$^0$ and $T$, and which is flat, and a "physical" one, $\gamma$, obtained by combining **g** and $t_\mathbf{x}$, and which is curved. The question then arises as to which metric we should refer the equations for the field of ether pressure, Eqs. (1) and (4). We assume that, due to the local character of these two equations, and for yet another reason, the *physical* space and time metrics must be used [2-3]. This means (i) that the gradient operator in Eq. (1), as well as the Laplace operator in Eq. (4), are given by

$$(\text{grad } \phi)^i = (\text{grad}_\mathbf{g} \phi)^i = g^{ij} \frac{\partial \phi}{\partial x^j}, \quad (g^{ij}) \equiv (g_{ij})^{-1}, \tag{8}$$

$$\Delta \phi = \Delta_\mathbf{g} \phi = \mathrm{div}_\mathbf{g}\, \mathrm{grad}_\mathbf{g}\, \phi = \frac{1}{\sqrt{g}} \frac{\partial}{\partial x^i}\left(\sqrt{g}\, g^{ij}\, \frac{\partial \phi}{\partial x^j}\right),\quad g \equiv \det(g_{ij}); \tag{9}$$

and (ii) that the time derivative in Eq. (4) must be understood as relative to the local time $t_\mathbf{x}$, that is

$$\frac{\partial \phi}{\partial t_\mathbf{x}} \equiv \frac{p_e^\infty}{p_e}\frac{\partial \phi}{\partial T} \equiv \frac{1}{\beta}\frac{\partial \phi}{\partial T}. \tag{10}$$

Now the "sound" velocity in the ether, $c_e$, depending on the local ether density $p_e$ as given by Eq. (2), must be assumed to be equal (everywhere and at any time) to the constant velocity of light $c$ [2]. This can be true only if $p_e = c^2 \rho_e + a$, and the constant $a$ must be zero since the pressure must cancel for nil density, thus we get: $p_e = c^2 \rho_e$. The modification of Eq. (4) accounting for SR and the equivalence principle is hence written as

$$\Delta p_e - \frac{1}{c^2}\frac{\partial^2 p_e}{\partial t_\mathbf{x}^2} = \frac{4\pi G}{c^2}\rho p_e. \tag{11}$$

In this equation, $\rho$ is now "the mass-energy density", a somewhat ambiguous expression to be precised in Section 5. Using assumption (A), one may rewrite Eq. (1) [now valid with the gradient defined in terms of the curved physical metric $\mathbf{g}$, Eq. (8)] and Eq. (11) in terms of the Euclidean metric $\mathbf{g}^0$ and the absolute time $T$ [4]:

$$\mathbf{g} = -\frac{c^2}{2}\,\mathrm{grad}_{\mathbf{g}^0} f = -\frac{c^2}{2}\,\mathrm{grad}_0 f,\quad f \equiv \beta^2 = (\gamma_{00})_\mathrm{E},\ (x^0 = cT), \tag{12}$$

$$\Delta_0 f - \frac{1}{f}\left(\frac{f_{,0}}{f}\right)_{,0} = \frac{8\pi G}{c^2}\rho,\quad \Delta_0 \equiv \Delta_{\mathbf{g}^0} \equiv \mathrm{div}_{\mathbf{g}^0}\,\mathrm{grad}_{\mathbf{g}^0},\ (x^0 = cT). \tag{13}$$

However, Eq. (13) is equivalent to Eq. (11) only if the "reference ether pressure" $p_e^\infty$ is time-independent, i.e., the universe is assumed "static at infinity", which may indeed be assumed, except for cosmological problems [4]. The simple form (12)-(13) of the main equations has important computational consequences.

## 4. The motion of a test particle as defined by an extension of Newton's second law

In the investigated theory, the motion of mass points and light-like particles, as well as the motion of any continuous medium (Sections 6 and 7), is governed by an extension of Newton's second law to curved space-time. The main reason to seek after such an extension was that Einstein's assumption, according to which free test particles follow space-time geodesics, gives a physical status to space-time. That the primary object of physics should be such mixture of space and time, is very difficult to accept for the "common sense", and has puzzling consequences such as the possibility of a travel back in time, with its well-known paradoxes. In contrast, if the Lorentz-Poincaré interpretation of SR is adopted, the space-time may be envisaged simply as a (very convenient and clever) mathematical tool. This is indeed the way in which Poincaré [31] introduced in 1905 the concept of the space-time as a 4-dimensional space with coordinates $x, y, z, ict$. According to the view adopted here, the "relativistic space-time couplings" are merely due to physical effects, on clocks and meters, of absolute motion and gravitation. Einstein's geodesic assumption could be compatible with this view only in the case that this assumption could be deduced from another one, directly compatible with the notion of distinct space and time. Such is Newton's assumption "Force = time-derivative of momentum".

In classical mechanics, it is indeed possible to rewrite Newton's second law, in a force field deriving from a *constant* potential, as the geodesic equation for a certain artificial space-time metric, and this *exactly* (*cf.* Mazilu [25]). In relativistic gravitation theory, it is known to be possible, conversely, to rewrite the spatial part of the geodesic equation as Newton's second law, again in the case of a *constant* gravitational field, i.e. for a time-independent space-time metric [23]. However, Landau & Lifchitz [23] did not actually prove that the obtained Newton law, thus three scalar equations, is equivalent to the geodesic equation, which involves four scalar equations. In the case of a variable metric, the various attempts to rewrite the geodesic equation with frame-dependent gravitational forces have been reviewed by Jantzen *et al.* [22]. As it appears from their review, and as it is also discussed in Ref. 5, none of these attempts is fully convincing. The reason is, in the author's opinion, that most of these researchers did not try to play the game "Newton's second law" with an *equivalent* equipment to that in classical mechanics, thus one reference frame (of which the possible changes may be discussed, if desired, in a *subsequent* step), with one spatial metric and one measure of physical time (Møller [27] did try this, but his pioneering proposal has one serious shortcoming, to be recalled hereafter). Instead, most authors retain the whole of the space-time metric with its ten independent components – in contrast with the seven scalars involved in the spatial metric (six) plus the local time (one).

To get a real, compelling equivalent of Newton's second law in a given reference frame, in the "relativistic" situation where the spatial metric **g** is curved by gravitation and *evolves with time*, and where clocks go differently at different places and times, is the objective that, the author believes, has been reached by the following proposal [4-5].

**i**) The precise expression of the *force* **F** will depend, of course, on each particular theory, and will be decomposed into a non-gravitational (non-universal) force $\mathbf{F}_0$, whose expression should be adapted from SR, and a gravitational force $\mathbf{F}_g$. The universal character of the latter, plus the assumption that SR must hold true locally, impose that it has the form $\mathbf{F}_g = m(v)\,\mathbf{g}$, where $m(v)$ is the relativistic inertial mass,

$$m(v) \equiv m(v=0).\gamma_v \equiv m(0).(1-v^2/c^2)^{-1/2}, \qquad (14)$$

the velocity vector **v** and its modulus $v$ being evaluated in terms of the local physical standards of space and time,

$$v^i \equiv dx^i/dt_\mathbf{x}, \qquad v \equiv [\mathbf{g}(\mathbf{v},\mathbf{v})]^{1/2} = (g_{ij}\,v^i\,v^j)^{1/2}, \qquad (15)$$

and where **g** is a space vector that should depend only on the current position and velocity of the test particle. The dependence on the velocity may be surprising at first if one remembers Newtonian gravity, but it means simply that the gravity acceleration **g** may contain an inertial part – as is to be expected if a general reference frame is considered. Even in a general reference frame, the data of the space-time metric determines uniquely the increment of the local time, $dt_\mathbf{x}$ ("synchronized" along the trajectory of the test particle, *cf.* Cattaneo [13] and Landau & Lifchitz [23]). This means that the ratio $dt_\mathbf{x}/d\chi$, with $\chi$ the parameter used to run along the trajectory, is unambiguously defined once a fixed reference frame is considered [5]. In the present theory, the preferred frame E is a globally synchronized reference frame (i.e., the $\gamma_{0i}$ components are zero, e.g. with the choice $x^0 = cT$ for the time coordinate). This implies that the trajectory may always be parametrized with the absolute time $T$ (which is also true for any other globally synchronized coordinate time, $T' = \phi(T)$), and that the relevant ratio is very simple, Eq. (6).

**ii**) There is no choice for the expression of the *momentum*, i.e. the product of the inertial mass and the velocity:

$$\mathbf{P} \equiv m(v)\,\mathbf{v}. \tag{16}$$

The difficult point is definition of the *time-derivative* of the momentum along the trajectory. Indeed, to define this consistently, we have to define, more generally, the derivative of any vector $\mathbf{w}(\chi)$ on the trajectory (defined by the dependence $x^i = x^i(\chi)$), and this trajectory is in a space (3-dimensional manifold) equipped with a *Riemannian* metric $\mathbf{g}$ that *varies* with the parameter $\chi$ along the trajectory. The variation referred to is not merely the spatial variation in $\mathbf{g}$ (or in its components $g_{ij}$) as the point $\mathbf{x} = (x^i)$ is changed (this variation is usual for a Riemannian manifold). It is also the time variation, precisely it is the fact that the spatial *field* $\mathbf{g}$ (the correspondence between *any* spatial point $\mathbf{x}'$ and the local metric $\mathbf{g}(\mathbf{x}')$), is not the same for different values of the parameter $\chi$ (which indeed plays the role of a time).

The first difficulty, i.e. the spatial variation of the metric, has been solved since a long time, and it is for this reason that the case of a constant gravitational field has been solved also: the correct time-derivative of a vector $\mathbf{w}(\chi)$ along a trajectory in a space equipped with a *fixed* Riemannian metric, is what is usually called the *absolute derivative* $D_0\mathbf{w}/D\chi$ (relative, however, to the considered metric $\mathbf{g}$ and to the considered parameter $\chi$; of course, if one changes $\chi$ to $\xi = \xi(\chi)$, one gets $D_0\mathbf{w}/D\xi = (d\chi/d\xi)D_0\mathbf{w}/D\chi$). The absolute derivative is usually introduced as a by-product of the "covariant derivative", the latter being valid for fields defined in volume domains of the manifold (e.g. Brillouin [9], Lichnerowicz [24]). However, the absolute derivative may be introduced specifically as the only consistent time-derivative of a vector along a trajectory, in the case of a fixed metric [2]. (Of course, the absolute derivative may be defined for manifolds with an arbitrary number of dimensions: e.g. in the case of the 4-dimensional space-time, the absolute derivative with respect to the space-time metric $\boldsymbol{\gamma}$ and to the parameter $s = c\tau$ with $\tau$ the proper time, $\Delta\mathbf{W}/\Delta s$, allows definition of the 4-acceleration of a test particle, $\mathbf{A} \equiv \Delta\mathbf{U}/\Delta s$ with $\mathbf{U} \equiv d\mathbf{x}/ds$ the 4-velocity.)

In order to define the time-derivative $D\mathbf{w}/D\chi$ of a vector $\mathbf{w}(\chi)$ along a trajectory $\mathbf{x}(\chi)$, in the case of a *variable* metric field $\mathbf{g} = \mathbf{g}_\chi$, we impose the following essential requirements [5]: 1) it must be a space vector depending linearly on $\mathbf{w}$; 2) it must reduce to the absolute derivative if $\mathbf{g}$ does not depend on $\chi$; 3) it must be expressed in terms of the space metric $\mathbf{g}$ and its derivatives; and 4) it must obey Leibniz' derivation rule for a scalar product, i.e.

$$\frac{d}{d\chi}\left(\mathbf{g}(\mathbf{w},\mathbf{z})\right) = \mathbf{g}\left(\mathbf{w}, \frac{D\mathbf{z}}{D\chi}\right) + \mathbf{g}\left(\frac{D\mathbf{w}}{D\chi}, \mathbf{z}\right). \tag{17}$$

It turns out that these requirements allow to define one and only one time-derivative, given by [5]:

$$D\mathbf{w}/D\chi \equiv D_0\mathbf{w}/D\chi + (1/2)\,\mathbf{t}.\mathbf{w}, \qquad \mathbf{t} \equiv \mathbf{g}_\chi^{-1}.\frac{\partial \mathbf{g}_\chi}{\partial \chi}. \tag{18}$$

In particular, Leibniz' rule (17) is responsible for the unique possible value of the coefficient multiplying the vector $\mathbf{t}.\mathbf{w} = (t^i{}_j w^j)$, i.e. $\lambda = 1/2$. In contrast, Møller's definition of Newton's second law [27] amounts to using the value $\lambda = 0$, i.e. to take the absolute time-derivative relative to the *frozen* space metric of the "time" $\chi_0$ where the derivative is to be calculated; thus, this time-derivative does *not* obey Leibniz' rule with the actual, variable metric, Eq. (17). Leibniz' rule is very important to obtain correct energy equations [4-5].

Now, to write Newton's second law in terms of local standards, the parameter $\chi$ must be the local time $t_\mathbf{x}$, "synchronized" along the trajectory. Moreover, one defines naturally for a mass particle the "purely material" energy (i.e., not accounting for the potential energy in the gravitational field [4]) as $E \equiv m(v)c^2$, so we finally write:

$$\mathbf{F}_0 + (E/c^2)\,\mathbf{g} = D\mathbf{P}/Dt_\mathbf{x}. \tag{19}$$

**iii**) For a light-like particle (photon, neutrino?), the energy $E$ of the particle (or rather the ratio $E/c^2$ i.e. the mass equivalent of the energy) plays the role of the inertial mass [13], so we define

$$\mathbf{P} \equiv (E/c^2)\,\mathbf{v}, \tag{20}$$

it being understood that $E$ is the usual, "purely material" energy, related to the frequency $\nu$ by Planck's relation

$$E = h\nu \tag{21}$$

(here $\nu$ is the frequency as measured with the local time, i.e. by the momentarily coincident clock). Equations (19) and (20) define one and the same Newton law for both mass particles and light-like test particles. Furthermore, the frequency $\nu$ deduced from $E$ by Eq. (21) is none other than de Broglie's frequency [30].

**iv**) Having thus defined a unique extension of Newton's second law, Eq. (19), to any space-time curved by gravitation, a question naturally arises: is this extended Newton law compatible with geodesic motion? Since geodesic motion applies only to *free* particles, we have to investigate the case $\mathbf{F}_0 = 0$ in Eq. (19); and since the gravity acceleration $\mathbf{g}$ has not been defined in the general statement of Newton's second law (although it is defined, in our theory, by Eq. (1) or equivalently by Eq. (12)), it is clear that the question can be reformulated thus: *which form of the gravity acceleration $\mathbf{g}$ is compatible with Einstein's geodesic motion?* Recall that Eq. (19) can be written in any reference frame (although the transformation law of $\mathbf{g}$ from one frame to another one has not been given, and does not seem to be easy to get: $\mathbf{g}$ is only assigned to be a spatial vector in a given reference frame). As in Newtonian theory, one may expect that the form of $\mathbf{g}$ should be simpler is some special class of reference frames. It turns out that the "convenient" frames are the "*globally synchronized* " ones, i.e. frames such that, in certain coordinates bound to the frame, the $\chi_{0i}$ components are zero. This is not surprising in this "neo-Newtonian" approach: in general frames, the time can be synchronized only on a trajectory, whereas the notion of a global simultaneity is essential (perhaps the most essential notion at all) in classical mechanics. According to Landau & Lifchitz [23], the very existence of globally synchronized frames is a property valid for generic space-times. However, there are "pathological" space-times in which it is not valid. The answer to the above question has been found [5] for such generic space-times: *In order that free particles follow space-time geodesics, it is necessary and sufficient that, in any globally synchronized reference frame, the gravity acceleration have the following expression*:

$$\mathbf{g}_{\text{geod}} = -c^2 \frac{\text{grad}_{\mathbf{g}} \beta}{\beta} - \frac{1}{2} \mathbf{g}^{-1} \cdot \frac{\partial \mathbf{g}}{\partial t_{\mathbf{x}}} \cdot \mathbf{v}, \quad \beta \equiv \sqrt{\gamma_{00}}. \tag{22}$$

**v)** It is seen that the "Einsteinian" gravity acceleration, Eq. (22), *differs* from the gravity acceleration assumed in the present theory by the second, velocity-dependent term in the right-hand side: this term is absent from Eq. (1), whereas, due to the equation $p_e = c^2 \rho_e$ (see after Eq. 10), the first term in the r.h.s. of Eq. (22) is just the r.h.s. of Eq. (1). Note that the velocity-dependent term cancels for a gravitational field that is constant in the considered frame. Therefore, another question arises: *how to phenomenologically characterize the assumed gravity acceleration, Eq. (1) with $p_e = c^2 \rho_e$?* The answer is [5]:

*Assume that in some "globally synchronized" reference frame* F *($\gamma_{0i} = 0$), the gravity acceleration be a space vector* **g** *depending only on the metric field* **γ**. *More precisely, assume that* **g** *does not depend on the time variation of* **γ** *and is linear with respect to the space variation of* **γ**. *In order that free particles follow space-time geodesics in the static case ($\gamma_{\mu\nu, 0} = 0$), it is necessary and sufficient that the general expression of vector* **g** *in the frame* F *be*

$$\mathbf{g} = -c^2 \frac{\text{grad}_{\mathbf{g}} \beta}{\beta} = -\frac{c^2}{2} \frac{\text{grad}_{\mathbf{g}} f}{f}, \quad f \equiv \gamma_{00} \equiv \beta^2, \tag{23}$$

*with* **g** *the space metric in* F.

This result characterizes the assumed field **g** independently of any heuristic consideration on the ether. The assumption that the metric field **γ** is a "spatial potential" for the field **g** is natural if one wants to account for the equivalence principle without going too far from Newtonian theory. Thus, Einstein's geodesic motion would be valid only in a static gravitational field.

## 5. Energy balance and energy conservation

We summarize the discussion in Ref. 4, where a few additional references are discussed. The concept of energy is an essential one in most of modern physics but, surprisingly, it can hardly be defined in modern gravitation theory, i.e. in general relativity (GR). Indeed, an extremely important feature of the energy is that it should be conserved, also locally, in the sense that a *local balance equation without any source term* should apply to the total energy. Now according to GR and other relativistic theories of gravitation based on general covariance plus Einstein's "equivalence principle" (in the standard form: "in a local freely falling frame, the laws of non-gravitational physics are the same as in SR"), the general "conservation equation" is the equation

$$(T^{\mu\nu}{}_{;\nu})_{\mu = 0, ..., 3} \equiv \mathbf{div}_{\boldsymbol{\gamma}} \mathbf{T} = 0 \tag{24}$$

for the energy-momentum tensor **T** [2]. But, as emphasized in most textbooks on GR, e.g. Landau & Lifchitz [23], this equation can not be considered as a true conservation equation (a balance equation without source term), for there is no Gauss theorem applying to the divergence of a second-order tensor in a curved Riemannian space. (In turn the main reason for this is that one can simply not define the integral of a vector field in a such space.) In more explicit terms: one cannot rewrite Eq. (24) in the form of a true conservation equation which would be valid for a generic coordinate system, i.e., which would be consistent with the principle of general covariance. One may, however, rewrite Eq. (24) in the form of a true conservation equation, if one accepts to restrict oneself to coordinate systems exchanging by *linear* coordinate transformations, as are Lorentz transformations in a flat space-time. Unfortunately, one may do this in many different ways, so that it is not clear which would be the correct way, even once one has specified the linear class of coordinate systems. Moreover, there is no reason to introduce such particular class, unless "the space-time has a particular symmetry", which means in fact that some *background metric* $\gamma^0$ on the space-time manifold, distinct from the physical metric $\gamma$, has some non-trivial group of isometries. The example of a such background metric that is relevant to the rewriting of Eq. (24) as a true conservation equation is that of a *flat* metric [4]: a such metric admits a particular class of coordinate systems, in which it reduces to the Galilean (Minkowskian) form. In summary, the search for a consistent concept of energy leads in GR to a contradiction with the very notion of "general relativity", since this search leads to restrict oneself to reference frames exchanging by Lorentz transformations of a flat background metric. In the present preferred-frame theory, on the other hand, we do have a flat background metric $\gamma^0$ and we do not even demand that the energy balance equation should be covariant by Lorentz transformations of this flat metric, so it would be hard to accept that the theory would not lead to a true conservation equation for the energy.

---

[2] Semi-colon means covariant derivative with respect to the space-time metric **γ**.

The obtainment of the energy equation for a mass point (pp. 42-43 in Ref. 4) is a modification of the elementary method used in classical mechanics to derive the energy equation in a force field deriving from a variable potential (the modifications are due to relativistic mechanics with a variable metric). Here, the assumed expression for **g** (Eq. 23)) derives from the potential $U' \equiv -c^2 \text{Log } \beta$. Thus, one evaluates the rate of work per unit rest mass for a "free" mass point,

$$dw/dt_\mathbf{x} \equiv \left(\mathbf{F}_g \cdot \mathbf{v}\right)/m_0 = \gamma_v \mathbf{g}.(d\mathbf{x}/dt_\mathbf{x}) = c\,\mathbf{g}.(d\mathbf{x}/ds) \qquad (25)$$

(where the point means scalar product **g**), using Newton's second law (19) (with $\mathbf{F}_0 = 0$), which involves the correct time-derivative (18). The result is

$$\frac{d}{dT}\left(c^2 \text{Log } \gamma_v\right) = \mathbf{g} \cdot \frac{d\mathbf{x}}{dT} = \left(\text{grad}_\mathbf{g}\, U'\right) \cdot \frac{d\mathbf{x}}{dT} = U'_{,i}\frac{dx^i}{dT} \equiv \frac{dU'}{dT} - \frac{\partial U'}{\partial T}. \qquad (26)$$

Using the definition of the potential $U'$, one rewrites this as

$$\frac{d(\gamma_v\,\beta)}{dT} = \gamma_v \frac{\partial \beta}{\partial T}, \qquad (27)$$

or, multiplying by $m_0 c^2$ with $m_0$ the rest mass:

$$\frac{d(E\beta)}{dT} = E\frac{\partial \beta}{\partial T}. \qquad (28)$$

Equation (28) shows clearly that the total energy of the mass point must be defined as

$$e_m \equiv E\beta = c^2 m_0 \gamma_v\,\beta, \qquad (29)$$

13which is a constant for a constant gravitational field. This is the total energy of the mass point, for it includes both its "purely material" energy $E$ (i.e. the energy equivalent of the relativistic inertial mass, thus including the "kinetic" energy) and its "potential" energy in the gravitational field, which may be defined as $e_{gm} = e_m − E = E (\beta − 1)$. (It is hence negative, as in NG.) It turns out that just the same equation (28) may be derived also for a light-like particle.

The deduction of the energy equation (28) from Eq. (27) is trivial, but it rests on the essential assumption that the rest mass $m_0$ of the free mass point is conserved in the motion (an assumption that is already used in the derivation of Eq. (26)). If we now consider a *dust*, i.e. a continuum made of coherently moving, non-interacting particles, each of which conserves its rest mass, we may apply Eq. (27) pointwise in the continuum. The conservation of the rest-mass of the continuum is most easily expressed in terms of the "background" Euclidean metric $\mathbf{g}^0$ and the associated volume measure $\delta V^0$, for it is then expressed as the usual continuity equation for the density of the rest-mass with respect to $\delta V^0$, which is $\rho_{00} \equiv \delta m_0/\delta V^0$. The density, with respect to $\delta V^0$, of the total energy of the dust, is $\varepsilon_m \equiv c^2 \rho_{00} \gamma_v \beta$, because

$$\delta e_m = c^2 \delta m_0\, \gamma_v \beta = c^2 \rho_{00}\, \delta V^0\, \gamma_v \beta = \varepsilon_m\, \delta V^0. \tag{30}$$

It turns out that $\varepsilon_m$ is none other than the $T^0{}_0$ component of the energy-momentum tensor $\mathbf{T}$ for the dust, whose $T^i{}_0$ components are $T^i{}_0 = T^0{}_0\, u^i/c$, with $u^i \equiv dx^i/dT$ the "absolute velocity" (with respect to the preferred frame, and evaluated with the absolute time $T$ ) [3]. Using this and rewriting Eq. (27) with the help of the continuity equation, one gets the *local balance equation for the continuum:*

$$cT^\mu{}_{0,\mu} \equiv c\left(\mathrm{div}_{\boldsymbol{\gamma}^0}\mathbf{T}\right)_0 = \frac{T^0{}_0}{\beta}\frac{\partial\beta}{\partial T}\quad (x^0 = cT), \tag{31}$$

where the identity applies when the spatial coordinates in the preferred frame are Cartesian (at least at the point considered), i.e. such that $g^0{}_{ij} = \delta_{ij}$ and $g^0{}_{ij,k} = 0$.

To rewrite this as a true conservation equation, we must use the field equation of the theory (Eq. (13)) so that the r.h.s. of Eq. (31), which in this form is a source term, be recast as a 4-divergence with respect to the flat metric $\boldsymbol{\gamma}^0$. In other words, we have to make the gravitational energy and its flux appear. To do this, we adapt the reasoning that leads to a conservation equation in NG: we observe that, due to Eqs. (12) and (13), one has [4]

$$\frac{8\pi G}{c^2}\rho f_{,0} = -\frac{2}{c^2}\mathrm{div}_0\left(f_{,0}\,\mathbf{g}\right) - \frac{1}{2}\left[\left(\frac{f_{,0}}{f}\right)^2 + \frac{4}{c^4}\mathbf{g}^2\right]_{,0},\quad \mathrm{div}_0 \equiv \mathrm{div}_{\mathbf{g}^0},\quad \mathbf{g}^2 \equiv \mathbf{g}^0(\mathbf{g},\mathbf{g}),\quad x^0 = cT. \tag{32}$$

On the other hand, using the condition $\gamma_{0i} = 0$, the r.h.s. of Eq. (31) may be rewritten as [4]

---

[3] In this paper, we shall use the energy units for tensor $\mathbf{T}$, whereas mass units were used in Refs. 4 and 7.
[4] Indices are raised and lowered with the help of the physical space-time metric $\boldsymbol{\gamma}$, unless explicitly mentioned otherwise.

$$\frac{T^0_{\ 0}}{\beta}\frac{\partial \beta}{\partial t} = \frac{T^0_{\ 0}}{2f}\frac{\partial f}{\partial t} = \frac{T^{00}}{2}\frac{\partial f}{\partial t} = \frac{c}{2}T^{00}f_{,0}. \tag{33}$$

Since $c^2\rho$ is "the mass-energy density" (in the preferred frame), it should be equal to $T^{00}$, or to $T^0_{\ 0}$, or still to $T_{00}$. But Eqs. (32) and (33) show that the source term on the r.h.s. of Eq. (31) can be rewritten as a flat 4-divergence, if and only if $c^2\rho = T^{00}$. Therefore, we must precisely define $c^2\rho$ as the $T^{00}$ component (in coordinates bound to the preferred frame, and such that $x^0 = cT$). This also means that the gravitational field reinforces itself, whereas, if we would assume $c^2\rho = T^0_{\ 0}$ or $c^2\rho = T_{00}$, the gravitational field would have a weakening effect on itself [4]. We therefore obtain the *local conservation equation for the continuum:*

$$\left(\mathrm{div}_{\gamma^0}\mathbf{T}\right)_0 + \frac{1}{8\pi G}\left\{\left[\mathbf{g}^2 + \frac{c^4}{4}\left(\frac{f_{,0}}{f}\right)^2\right]_{,0} + \mathrm{div}_0\left(c^2 f_{,0}\,\mathbf{g}\right)\right\} = 0 \quad (x^0 = cT). \tag{34}$$

Although this equation has been derived for a dust, we assume that it holds true with **T** the *total* energy-momentum tensor of any kind of continuum (involving material particles and/or non-gravitational fields). This assumption is justified by the mass-energy equivalence and the universality of the gravitation force. Equation (34) rules the exchange between the total energy of matter, whose density is given by $\varepsilon_m = T^0_{\ 0}$ (cf. Eq. (30)), and the purely gravitational energy, whose density $\varepsilon_g$ is defined by

$$\varepsilon_g = \frac{1}{8\pi G}\left[\mathbf{g}^2 + \frac{c^4}{4}\left(\frac{f_{,0}}{f}\right)^2\right]. \tag{35}$$

This exchange occurs through the intermediate of the flux of the total matter energy, defined as the space vector with components $cT^i_{\ 0}$, and the flux of the gravitational energy, defined as the space vector $c^3 f_{,0}\,\mathbf{g}/(8\pi G)$. Note that the total energy of matter contains also the negative potential energy of matter (and/or non-gravitational fields) in the gravitational field: for a dust, $T^0_{\ 0} = \varepsilon_m$ is the density of the total energy of the individual particles, defined by $e_m = c^2 m_0 \gamma_v \beta$ (cf. Eqs. (29) and (30)). Of course, the local conservation of the energy implies a global conservation (of the global amount of *total energy*, i.e. gravitational energy plus total energy of matter), under asymptotic conditions ensuring that the global energy is finite [4].

### 6. Continuum dynamics and matter creation/destruction

By the foregoing induction (from a dust to a general continuum), we have got one scalar local equation for a continuous medium, i.e. the energy conservation, which, for a general continuum, is *substituted for the mass conservation*. However, for a point particle, we have four equations: the three equations of motion (19), plus the conservation of the rest-mass, and it is easy to convince oneself that one also needs exactly four independent dynamical equations for a continuum, in addition to the state equation. For instance, there are indeed four dynamical equations in classical continuum mechanics: Newton's second law plus the continuity equation. The same number applies also in GR, where the dynamical equations make the well-known 4-vector equation (24).

In order to get the required four scalar "equations of motion" for a continuous medium, we may again use the general principle of induction from a dust to a general behaviour, once the equation for dust has been expressed in terms of the energy-momentum tensor **T**. The most expedient way to operate this principle turns out to be passing through the expression of the *4-acceleration* [7]. The latter expression has been obtained for a free particle [5]. (The spatial components of the 4-acceleration and its time component were deduced from Newton's second law (19) and the energy equation (27), respectively, by using the relation between the Christoffel symbols of the spatial metric and those of the space-time metric.) It happens to be simpler in covector form [4]:

$$A_0 = \frac{1}{2} g_{jk,0} U^j U^k, \quad A_i = -\frac{1}{2} g_{ik,0} U^0 U^k. \tag{36}$$

(By the way, Eq. (36) shows at once that, in the present theory, Einstein's geodesic motion is recovered only for a gravitational field that is constant in the preferred frame: **A** = 0 is true for whatever 4-velocity **U** if and only if $g_{ij,0} = 0$.) For a dust, the **T** tensor has the form $T^{\mu\nu} = c^2 \rho^* U^\mu U^\nu$ with $\rho^*$ the proper rest-mass density, and, for any material continuum, the 4-acceleration may be expressed as $A^\mu = U^\nu U^\mu_{;\nu}$. Using this and the mass conservation, assumed valid by definition for a dust, one may rewrite Eq. (36), for a dust, as

$$T_\mu{}^\nu{}_{;\nu} \equiv T^\nu{}_{\mu;\nu} = b_\mu, \tag{37}$$

where $b_\mu$ is defined by

$$b_0(\mathbf{T}) \equiv \frac{1}{2} g_{jk,0} T^{jk}, \quad b_i(\mathbf{T}) \equiv -\frac{1}{2} g_{ik,0} T^{0k}. \tag{38}$$

Now the induction principle means simply that *Eq. (37) is the general equation for continuum dynamics* in the present theory. It thus plays the role played in GR by Eq. (24). The r.h.s. of Eq. (37), $b_\mu$ as defined by Eq. (38), is a 4-covector for transformations of the group (7), and so is also Eq. (37). The time component of Eq. (37) ($\mu = 0$) is equivalent to the energy balance equation (31), hence also to the energy conservation equation (34) [7]. Moreover, for a dust, i.e. the case $T^{\mu\nu} = c^2 \rho^* U^\mu U^\nu$, Eq. (37) *implies* the mass conservation, i.e.

$$\left(\rho^* U^\nu\right)_{;\nu} = 0, \tag{39}$$

and (again for a dust), Eq. (37) *implies* also the expression (36) for the 4-acceleration, which is characteristic for free motion in the present theory, i.e. $\mathbf{F}_0 = 0$ in Newton's second law (19) [7]. Thus, Eq. (37) plus the relevant definition of the energy-momentum tensor as a function of the state variables (which, for dust, consist of the single variable $\rho^*$) characterize completely the dynamical behaviour of dust. This is an important consistency test.

A general continuum may thus be phenomenologically defined by the expression of tensor **T** as a function of some state variables, and Eq. (37) determines how the continuum couples to gravitation in the present theory (of course, Eq. (37) reduces to the equation valid in SR, i.e., $T^{\mu\nu}{}_{,\nu} = 0$, if there is no gravitational field). But, for a general continuum, the *energy* is conserved, and this is in general *incompatible with the exact mass conservation:* in the case of a variable gravitational field, there are exchanges between the gravitational energy and the total energy of matter, so one may a priori expect that, in general, the rest-mass will not be conserved − except for the special case of a dust. One may already verify this for the simple case of a perfect isentropic fluid. The energy-momentum tensor of a perfect fluid is in general

$$T_{\text{fluid}}{}^{\mu\nu} = (\mu^* + p)\, U^\mu U^\nu - p\, \gamma^{\mu\nu}, \qquad (40)$$

with $p$ the pressure and $\mu^*$ the volume density of the rest-mass plus internal energy in the proper frame: in energy units,

$$\mu^* \equiv \rho^*(c^2 + \Pi), \qquad (41)$$

where $\Pi$ is the internal energy per unit rest mass. For a perfect fluid, the isentropy condition is simply

$$d\Pi + p\, d(1/\rho^*) = 0. \qquad (42)$$

For instance, for a barotropic fluid, one assumes $\mu^* = \mu^*(p)$; then, $\rho^*$ and $\Pi$ also depend on $p$ only, $\Pi$ being given by [18]

$$\Pi(p) \equiv \int_0^p \frac{dq}{\rho^*(q)} - \frac{p}{\rho^*(p)}. \qquad (43)$$

Due to Eq. (43), a barotropic fluid is automatically isentropic.

For any isentropic fluid, Eq. (37) leads to the following *equation for mass creation/destruction:*

$$\left(\rho^* U^\nu\right)_{;\nu} \left(1 + \frac{\Pi + p/\rho^*}{c^2}\right) = -\frac{pU^0}{2c^2} \frac{f_{,0}}{f}, \qquad (44)$$

which indeed shows that, except for the limit case of a dust ($p = 0$) and for the limit case of a constant gravitational field ($f_{,0} = 0$), the rest-mass is not conserved − according to the present theory [7]. In contrast, GR and other relativistic theories are based on Eq. (24), which, for an isentropic fluid, implies the conservation of the rest-mass, Eq. (39) (*cf.* Chandrasekhar [15]). On the contrary, the present theory predicts that matter may really be produced or destroyed, due to the variation of the gravitational field. Prigogine *et al.* [32] consider that matter should be produced by a such exchange (albeit in an irreversible way, excluding matter *destruction*), and this exchange would indeed seem *a priori* natural in a theory with conserved energy, due to the mass-energy equivalence. However, matter production can hardly happen so in GR: we insist that, due to the equation for continuum dynamics that goes with geodesic motion, i.e., $T^{\mu\nu}{}_{;\nu} = 0$, matter can only be produced if one phenomenologically inserts an additional term, which is *not* determined by the set of the state equation plus the Hilbert-Einstein equations [12, 19, 32]. Roughly speaking, this "creation term" appears thus as an *ad hoc*, adjustable source term, which is used to allow production of matter in some cosmological models. It seems interesting to investigate the possibility that matter might be produced (*or destroyed*) by an exchange with the gravitational field (a more complete discussion of this question is given in Ref. 7, that includes in particular a discussion of the thermodynamical constraints). Yet in our opinion, such exchange should not be considered merely in a cosmological context, but actually for any gravitational field.

Precisely, the way in which matter production occurs in any variable gravitational field, as implied by Eq. (44), may seem dangerous for the present theory, because it would mean that matter is continuously produced or wasted away under our eyes. However, the rates would be extremely small and often the mean gain would be rather close to zero [7]. If the absolute velocity **V** of the solar system is of the order 300 km/s, the main contribution in Eq. (44) would come from that variation of the gravitational field which is due to the translation of any celestial body through the ether, giving a creation rate (the amount produced per unit time in some material domain, divided by the mass of that domain)

$$C \equiv \frac{\hat{\rho}}{\rho} \approx \frac{p}{c^2 \rho} \frac{GM}{c^2 R} \frac{r}{R} \frac{V_r}{R} \qquad (V_r \equiv \mathbf{V} \cdot \mathbf{e}_r), \qquad (45)$$

with $M$ and $R$ the mass and radius of the spherical celestial body, whose attraction $g$ ($g = GM/r^2$ outside the body) dominates in its near environment ($\mathbf{e}_r$ is the unit radial vector). At a fixed point on the surface of a body in self-rotation (as it is indeed the case for the planets), the corresponding contribution would be exactly cyclic at the equator, and instead would constantly accumulate production of matter (resp. destruction) at one pole (resp. at the other pole). Near the surface of the Earth, for instance, the ratio $p/(c^2\rho)$ can hardly take values much higher than $10^{-12}$ (which is its value in the air at the atmospherical pressure). A ratio equal to $10^{-12}$ leads to a maximum value of the creation rate $C_{max} \approx 3.\,10^{-23}\,\mathrm{s}^{-1}$ (obtained for $\mathbf{e}_r$ parallel to $\mathbf{V}$, and with $V \approx 300$ km/s), which seems very difficult to detect [7]. Hence, it might be the case that this new form of energy exchange be a real phenomenon. Needless to emphasize, this would be interesting.

## 7. Gravitationally-modified Maxwell equations and their link to photon trajectories

As emphasized by Will [40], any viable theory of gravitation should be "complete", in the sense that it should match with the whole of physics. However, even GR does not match with quantum physics, because a consistent theory of "general relativistic quantum gravity" does not exist yet, in fact it is not sure that such theory is still expected seriously. Hence, accounting for the present state of physics, a reasonable demand of completeness is that, in addition to a mechanics, an alternative theory of gravitation should involve also a description of the electromagnetism in the presence of gravitation [40]. This description should lead to a set of "gravitationally-modified" Maxwell equations, and a further requirement is that these modified equations should be consistent with the geometrical optics of the theory as defined by the trajectories of light-like particles. In other words, the theory should tell us how the electromagnetic *rays* may be defined from some particular electromagnetic *waves*, and it should prove that these rays do move as light-like particles do it under the gravitation. In GR and other metric theories, the modification of the Maxwell equations is rather easy, because it is an application of the well-known rule "comma goes to semi-colon": due to the commonly accepted formulation of Einstein's equivalence principle, one just has to substitute the curved physical metric $\gamma$ for the flat one into the covariant expression of any law of non-gravitational physics. (However, this rule is generally ambiguous when the expression contains higher-order derivatives. So in the case of Maxwell's equations, one is carefully warned not to apply the rule to the equations for the potential, and instead to apply the rule to the equations for the field tensor **F**.) Further, in GR and other metric theories, the transition from wave optics to geometrical optics is most rigorously made with the help of the discontinuities equations for an electromagnetic shock wave, with the result that the electromagnetic rays are the bicharacteristics of Maxwell's equations, corresponding to "null" fields, and following light-like geodesics (*cf.* e.g. Synge [37]).

In the present theory, the electromagnetic field tensor **F** is stated to derive from a 4-potential in the usual way: in particular it is antisymmetric, $F_{\mu\nu} = -F_{\nu\mu}$, and the first group of the Maxwell equations is satisfied,

$$F_{\lambda\mu,\nu} + F_{\mu\nu,\lambda} + F_{\nu\lambda,\mu} = F_{\lambda\mu;\nu} + F_{\mu\nu;\lambda} + F_{\nu\lambda;\mu} = 0. \qquad (46)$$

However, we cannot use the "comma goes to semi-colon" rule, because this theory is not a metric theory (although it does endow the space-time with a curved physical metric **γ**). Specifically, the form of the second group of Maxwell's equations in metric theories,

$$F^{\mu\nu}{}_{;\nu} = -4\pi \frac{J^\mu}{c} \qquad (47)$$

(where $J^\mu$ is the 4-current), implies that, *in vacuo*, the energy-momentum tensor of the electromagnetic field,

$$T_{\text{field}}{}^{\mu\nu} \equiv (-F^\mu{}_\lambda F^{\nu\lambda} + \frac{1}{4}\gamma^{\mu\nu} F_{\lambda\rho} F^{\lambda\rho})/4\pi, \qquad (48)$$

obeys Eq. (24), instead of Eq. (37) – and Eqs. (24) and (37) are equivalent only for a constant gravitational field. In the proposed theory, the Maxwell equations are obtained as an application of Newton's second law for a charged dust subjected to the gravitation force and to the Lorentz force. Hence, we begin with a short outline of continuum dynamics in the presence of gravitational and non-gravitational forces [8]. Just the same method of induction from a dust to a general continuum is used as in Sect. 6, the difference being that now the dust particles are subjected, in addition, to a non-gravitational *external* force. The density **f** of the latter is assumed given, and is expressed in terms of the physical volume measure $\delta V = \sqrt{g}\, dx^1\, dx^2\, dx^3$ (with $g \equiv \det(g_{ij})$). Thus, the non-gravitational external force over a volume element of the dust is

$$\delta \mathbf{F}_0 = \mathbf{f}\, \delta V. \qquad (49)$$

(In accordance with the spirit of field theory, the particles are still assumed to not interact directly: instead, their presence produces a field that exerts a force on them. This is more intuitive with the concept of a space-filling fluid ether, of course.) Adapting the energy equation (28) and then the expression (36) of the 4-acceleration for an individual particle to the situation with a non-gravitational force, and using the same method as in Sect. 6, one gets

$$T^{0\nu}{}_{;\nu} = b^0(\mathbf{T}) + \frac{\mathbf{f}.\mathbf{v}}{c\beta}, \qquad T^{i\nu}{}_{;\nu} = b^i(\mathbf{T}) + f^i, \qquad (50)$$

where $b^\mu (\mathbf{T})$ is the vector deduced from the covector (38) by raising up the indices with $\boldsymbol{\gamma}$.

The expression of the Lorentz force on a point particle with charge $q$ may be derived uniquely from the requirements that (i) it must be an invariant space vector by the group (7), and (ii) this vector must reduce to the classical expression in the absence of gravitation (in Galilean coordinates for the flat metric $\boldsymbol{\gamma} = \boldsymbol{\gamma}^0$),

$$F^i \equiv \frac{d}{dt}\left(m_0 \gamma_v \frac{dx^i}{dt}\right) = q\left(F^i{}_0 + F^i{}_j \frac{v^j}{c}\right). \tag{51}$$

By this method, the expression of the Lorentz force in the presence of gravitation is found to be [8]

$$F^i = q\left(\frac{F^i{}_0}{\beta} + F^i{}_j \frac{v^j}{c}\right) = \frac{q}{\gamma_v} F^i{}_\mu U^\mu. \tag{52}$$

Considering then a continuous medium (a "charged dust"), we define $\rho_{el} \equiv \delta q/\delta V$ and $J^\mu \equiv \rho_{el} \, dx^\mu/dt_\mathbf{x}$. The density of the Lorentz force is written, in accordance with Eqs. (49) and (52), as

$$f^i \equiv \frac{\delta F^i}{\delta V} = F^i{}_\mu \frac{J^\mu}{c}. \tag{53}$$

The charged dust obeys the equation for continuum dynamics in the presence of gravitation and the Lorentz force, Eq. (50) with $\mathbf{f}$ given by Eq. (53). On the other hand, the *total* energy-momentum is the sum $\mathbf{T} = \mathbf{T}_{dust} + \mathbf{T}_{field}$, and this total tensor must obey the equation for continuum dynamics in the presence of gravitation and without any non-gravitational external force, Eq. (37). Due to the linearity of the expression (38) of $b^\mu$ as a function of $\mathbf{T}$, it follows that the energy-momentum tensor of the electromagnetic field must satisfy

$$T_{field}{}^{0\nu}{}_{;\nu} = b^0(\mathbf{T}_{field}) - \frac{\mathbf{f}.\mathbf{v}}{c\beta}, \quad T_{field}{}^{i\nu}{}_{;\nu} = b^i(\mathbf{T}_{field}) - f^i. \tag{54}$$

In words: *the electromagnetic field may be considered as a "material" continuum subjected to the gravitation and to the opposite of the Lorentz force. This gives the gravitational modification of Maxwell's second group in the present theory*. Indeed, Eq. (54) may be rewritten as

$$F^\mu{}_\lambda F^{\lambda\nu}{}_{;\nu} = 4\pi b^\mu(\mathbf{T}_{field}) - 4\pi F^\mu{}_\lambda \frac{J^\lambda}{c} \tag{55}$$

and this, in the case of an invertible 4-4 matrix $(F^\mu{}_\nu)$, may in turn be written in the form

$$F^{\mu\nu}{}_{;\nu} = 4\pi\left(G^{\mu}{}_{\nu} b^{\nu}(\mathbf{T}_{\text{field}}) - \frac{J^{\mu}}{c}\right), \qquad \mathbf{G} \equiv \mathbf{F}^{-1} \quad . \tag{56}$$

Equation (56) is what corresponds, in the present theory, to Eq. (47) of GR, and *it reduces to Eq. (47) for the case of a constant gravitational field* (in the preferred frame). The condition det $\mathbf{F} \neq 0$ is equivalent to $\mathbf{E}.\mathbf{B} \equiv -\varepsilon^{\mu\nu\rho\psi} F_{\mu\nu} F_{\rho\psi}/8 \neq 0$, it is hence satisfied by "generic" electromagnetic fields. However, it is not satisfied by the simplest examples of e.m. fields, i.e. purely electric fields, purely magnetic fields, and "simple" electromagnetic waves (also called "null fields"), for which fields one may use only the weaker equation (55). A striking consequence of Eq. (56) is that *the electric charge is not conserved in a variable gravitation field,* since Eq. (56) implies that

$$\hat{\rho}_{\text{el}} \equiv \left(J^{\mu}\right)_{;\mu} = c\left(G^{\mu}{}_{\nu} b^{\nu}(\mathbf{T}_{\text{field}})\right)_{;\mu} . \tag{57}$$

However, it is difficult to give estimates without having recourse to a numerical work (in contrast to the situation for matter production, Sect. 6), hence this interesting and dangerous prediction is yet to be precised.

Equation (55) is sufficient to make the transition from wave optics to geometrical optics in the theory. The geometrical optics *in vacuo* is defined, in the present theory, as the study of the trajectories of light-like particles, which are ruled by the proposed extension (19) of Newton's second law, without any non-gravitational force, thus $\mathbf{F}_0 = 0$ in Eq. (19). As we have seen, the modified second group of the Maxwell equations, Eq. (55) or Eq. (56), is just the expression of Newton's second law for that continuous medium which is defined by the energy-momentum tensor of the electromagnetic field. *In vacuo*, a volume element of the field continuum will be subjected to the gravitation *and*, in general, to *internal forces* exerted by the neighbouring elements. Therefore, the condition $\mathbf{F}_0 = 0$, which must apply to an isolated photon, has to be replaced by the double condition that (i) the field continuum is subjected to zero non-gravitational external force, i.e., $\mathbf{f} = 0$ in Eq. (50), and (ii) it is subjected to zero internal force. Condition (ii) means that the continuum behaves like a dust (a dust of photons), i.e., that its energy-momentum tensor has the form

$$T^{\mu\nu} = V^{\mu} V^{\nu}. \tag{58}$$

One may indeed show that, independently of the exact physical nature of the material or field, Eq. (58) is the necessary and sufficient condition ensuring that any volume element of the continuum is subjected merely to the gravitation and to the non-gravitational *external* force density $\mathbf{f}$, thus with zero *internal* force [8]: in that case, the spatial part of the dynamical equation of the theory, Eq. (50)$_2$, is equivalent to

$$\mathbf{f}\,\delta V + \frac{\delta E}{c^2}\,\mathbf{g} = \frac{D}{Dt_{\mathbf{x}}}\left(\frac{\delta E}{c^2}\mathbf{v}\right), \qquad \mathbf{v} = \frac{1}{\beta}\mathbf{u}\,, \tag{59}$$

where the purely material energy of the element is defined, consistently with Eqs. (29) and (30), as

$$\delta E = \delta e_{\text{m}}/\beta = T^0{}_0\,\delta V^0/\beta = T^0{}_0\,\delta V, \tag{60}$$

and where the absolute velocity of the element, $\mathbf{u} = d\mathbf{x}/dT$, is defined by

$$u^i \equiv c\, T^i{}_0 / T^0{}_0, \tag{61}$$

which is the velocity of the flux of the total matter energy, as Eq. (31) shows.

Now, in the case of the energy-momentum tensor of the electromagnetic field (Eq. (48)), condition (ii) turns out to be equivalent to saying that the e.m. field is a null field (with both invariants equal to zero), whereas, due to Eqs. (53) and (54), condition (i) is equivalent to $J^\mu = 0$, i.e. to the requirement that an "empty" domain is considered (i.e. a domain in which the e.m. field is the only form of "matter"). Thus, *a "dust of free photons" is exactly a null field in vacuo*. Furthermore, for a such field, the second group of Maxwell's equations is Eq. (55) with $J^\mu = 0$, which is equivalent to Eq. (54) with $\mathbf{f} = 0$, hence this group is equivalent to the dynamical equation for the electromagnetic dust *in vacuo*:

$$T_{\text{field}}{}^{0\nu}{}_{;\nu} = b^0(\mathbf{T}_{\text{field}})\ ,\ T_{\text{field}}{}^{i\nu}{}_{;\nu} = b^i(\mathbf{T}_{\text{field}}). \tag{62}$$

As we have recalled above, Eq. (62)$_2$ may be rewritten, for a null field, as

$$\frac{\delta E_{\text{field}}}{c^2}\mathbf{g} = \frac{D}{Dt_\mathbf{x}}\left(\frac{\delta E_{\text{field}}}{c^2}\mathbf{v}_{\text{field}}\right), \qquad \mathbf{v}_{\text{field}} = \frac{1}{\beta}\mathbf{u}_{\text{field}}, \tag{63}$$

where $\mathbf{u}_{\text{field}}$ is defined by Eq. (61), thus $u_{\text{field}}{}^i \equiv c\, T_{\text{field}}{}^i{}_0 / T_{\text{field}}{}^0{}_0$ [and with $\delta E_{\text{field}} \equiv T_{\text{field}}{}^0{}_0\, \delta V$, Eq. (60)]. This definition of $\mathbf{u}_{\text{field}}$, together with Eq. (58) and the fact that $\mathbf{T}_{\text{field}}$ has zero trace, imply that $|\mathbf{v}_{\text{field}}| = c$ [8]. This and Eq. (63) allow us to conclude: *in the case of a null field in vacuo, each trajec-tory of the e.m. energy flux is a photon trajectory of the present theory*, as defined by Eq. (19) for a free light-like particle ($\mathbf{F}_0 = 0$). This is the link between wave optics and ray optics in the proposed theory.

## 8. Consistency with observations and the question of the preferred-frame effects

This tentative ether theory is now rather complete, and it is self-consistent. The obvious question is: does this theory agree with experiment? There is a vast amount of experimental and observational data as regards gravitational physics [40], so that a very detailed analysis should be performed, of course. For instance, there are numerous tests of the "*weak equivalence principle*", but the latter is none other, after all, than the statement that gravitation is a universal force. Due to the basic equation of motion in the theory, i.e. "Newton's second law" (19), this is obviously true in this theory. There are also tests of the more specific statement that: "in a local freely falling frame, the laws of non-gravitational physics are the same as in SR", which is *Einstein's equivalence principle* (EEP), and which we also call the equivalence principle in the standard form, because a different equivalence principle is postulated in the present theory (Sect. 3). It should be clear that EEP is *not* true in this theory, since EEP implies Eq. (24) for continuum dynamics, as opposed to Eq. (37). Let us recall, however, that these two equations are equivalent for the case of a constant gravitational field, and note that the best-known theoretical frame for testing EEP, known as the *THεμ* formalism, is restricted to static gravitational fields [40], and so does not allow to analyse experiments that should decide between EEP and the proposed equivalence principle. Furthermore, as well as for any theory based on EEP and Eq. (24), our Equation (37) and the whole theory are in full agreement with the assumption of a *universal coupling*, since the same equation applies to any kind of matter and/or non-gravitational field. For these two reasons, it would be probably quite difficult to find laboratory experiments accurate enough to distinguish between the usual form of the equivalence principle and the alternative form which is postulated in the present theory.

Hence, we are inclined to believe that the main challenge for this preferred-frame theory is to recover the "classical tests" of GR, i.e. the effects of gravitation on light rays and the general relativistic corrections to Newtonian celestial mechanics. The latter consist essentially in the prediction, by GR, of Mercury's very small residual advance in perihelion but, in our opinion, one should pose the question in a more general way: does the theory produce a celestial mechanics which is more accurate than Newton's theory?

In order to investigate the effects of gravitation on light rays and the corrections made by this non-linear theory to Newtonian celestial mechanics, it is necessary, as well as in GR, to develop an iterative approximation scheme, i.e., to develop a *post-Newtonian (pN) approximation scheme*. The pN approximation *scheme* is the method of asymptotic expansion of the dependent variables and the equations in powers of a small parameter $\varepsilon$, which is defined by $U_{max}/c^2 \equiv \varepsilon^2$, with $U_{max}$ the maximum value of the Newtonian potential in the considered gravitating system (assumed isolated, and the gravitational field being assumed weak and slowly varying; *cf.* Fock [18], Chandrasekhar [14], Weinberg [38], Misner *et al.* [26], Will [40]). (The term *pN approximation* alone usually makes reference to the approximation immediately following the first, "Newtonian" approximation; this second approximation is largely sufficient in the solar system.) Actually, the small parameter will be taken simply as $\varepsilon' = 1/c$ as in Refs. 14 and 18. Note that, choosing the units such that $U_{max} = 1$, we get indeed $\varepsilon = \varepsilon'$. More generally, constraining the units merely so that $U_{max} \approx 1$, we may take $\varepsilon' = 1/c$ as the small parameter. Then all relevant quantities such as $U$, $v$, etc., are O(1). Choosing the time coordinate as $x^0 = T$ (instead of $cT$), the assumption of a slowly varying gravitational field is then automatically satisfied. Moreover, only one term among two successive ones appears in the relevant expansions. Whereas the usual explanation makes appeal to the behaviour under time reversal [26, 40], we note that, in the proposed ether theory, all non-Newtonian effects come from the "ether compressibility", $K \equiv 1/c^2$. So $K$ itself (or $U_{max} K$, if the units are not constrained so that $U_{max} \approx 1$) could be considered as the small parameter, whence the appearance of only one among two successive terms in any expansion with respect to $1/c = \sqrt{K}$ – the leading ("Newtonian") term giving the parity. Finally, in a preferred-frame theory, the absolute velocity $V$ of the mass-center of the system with respect to the ether frame should not exceed the order $\varepsilon c$, as is the case for the typical orbital velocity $v$ in the mass-center frame. We note that, if $V$ is approximately 300 km/s for the solar system (as one finds if one assumes that the cosmic microwave background is "at rest" with respect to the preferred frame [40]), then one has indeed $V/c \leq \varepsilon$ in the solar system [6], because there $\varepsilon^2 \equiv U_{max}/c^2 \approx 10^{-5}$ [26].

**i**) *Expansion of the metric and the field equation in the preferred frame*
The leading expansion is that of the scalar field, $\beta$ or $f = \beta^2$ :

$$\beta = 1 - U/c^2 + S/c^4 + ... , \qquad f = 1 - 2U/c^2 + (U^2 + 2S)/c^4 + ... = 1 - 2U/c^2 + A/c^4 + ... \quad (64)$$

The space metric deduced from the Euclidean metric $\mathbf{g}^0$ by assumption (A) (Sect. 3) is then obtained [6] as

$$g_{ij} = g^0_{ij} + (2U/c^2)h^{(1)}_{ij} + O(1/c^4), \qquad h^{(1)}_{ij} \equiv (U_{,i}\, U_{,j})/(g^{0\,kl}\, U_{,k}\, U_{,l}) \tag{65}$$

(with $g^0_{ij} = g^{0\,ij} = \delta_{ij}$ in Cartesian coordinates), and the space-time metric is

$$\gamma_{00} = c^2 f = c^2(1 - 2U/c^2 + A/c^4 + ...), \quad \gamma_{0i} = 0, \; \gamma_{ij} = -g_{ij}. \tag{66}$$

The mass-energy density $\rho = [(T^{00})_E]/c^2$ may be written in the form

$$\rho = \rho^0 + w^1/c^2 + ..., \tag{67}$$

where $\rho^0$ is the conserved mass density which is found at the first approximation (expanding, for a perfect fluid, the energy equation (34), one finds that $\rho^0$ obeys the usual continuity equation and that mass is conserved at the pN approximation also). The pN expansion of the field equation (13) follows easily from Eqs. (64) and (67):

$$\Delta_0 U = -4\pi G \rho^0, \tag{68}$$

$$\Delta_0 S = 4\pi G w^1 - \Delta_0 U^2/2 - \partial^2 U/\partial t^2, \quad \text{or} \quad \Delta_0 A = 8\pi G w^1 - 2\partial^2 U/\partial t^2. \tag{69}$$

**ii**) *Post-Newtonian equations of motion for a test particle in the preferred frame*

Using the energy equation (28), one first rewrites Newton's second law for a free test particle, Eq. (19) with $\mathbf{F}_0 = 0$, as [6]

$$\frac{du^i}{dT} = -\Gamma'^i_{00} - \Gamma'^i_{0j} u^j - \Gamma^i_{jk} u^j u^k + \left(\Gamma'^0_{00} + 2\Gamma'^0_{0j} u^j\right) u^i \quad (x^0 = T), \tag{70}$$

where the $\Gamma'^\mu_{\nu\rho}$ symbols are the Christoffel symbols of the space-time metric and the $\Gamma^i_{jk}$'s are those of the space metric (with $\Gamma^i_{jk} = \Gamma'^i_{jk}$ for the spatial indices, due to the fact that $\gamma_{0i} = 0$). Therefore, expanding the equation of motion amounts to expanding the Christoffel symbols, using Eqs. (64)-(66). In doing so at the pN level, one has to distinguish between the case of a photon, for which the velocity $\mathbf{u}$ is $O(c)$, and the case of a mass particle moving under the action of the weak gravitational field, for which $\mathbf{u}$ is $O(1)$. Hence, the expanded equation of motion involves less terms for a photon, for which it is, in Cartesian coordinates for metric $\mathbf{g}^0$:

$$\frac{du^i}{dT} = U_{,i} - 2\left(U_{,j} u^j\right)\frac{u^i}{c^2} - \left((Uh^{(1)}_{ij})_{,k} + (Uh^{(1)}_{ik})_{,j} - (Uh^{(1)}_{jk})_{,i}\right)\frac{u^j u^k}{c^2} + O\left(\frac{1}{c}\right). \tag{71}$$

An important point is that Eq. (71), derived from Newton's second law, is nevertheless *undistinguishable from the pN expansion of the equation for null space-time geodesics*. For a mass point, the expanded equation is

$$\frac{du^i}{dT} = U_{,i} - \frac{1}{c^2}\left[\frac{A_{,i}}{2} + 2Uh^{(1)}{}_{ij}U_{,j} + \left(Uh^{(1)}{}_{ij}\right)_{,0} u^j + u^i\left(U_{,0} + 2U_{,j}u^j\right)\right]$$

$$-\left((Uh^{(1)}{}_{ij})_{,k} + (Uh^{(1)}{}_{ik})_{,j} - (Uh^{(1)}{}_{jk})_{,i}\right)\frac{u^j u^k}{c^2} + O\left(\frac{1}{c^4}\right). \quad (72)$$

Note that, in the pN equation of motion for a photon, Eq. (71), the "Newtonian" gravity acceleration $\mathbf{g}^0 \equiv \text{grad}_0\, U$ (with components $U_{,i}$ in the Cartesian coordinates utilized) intervenes at the same order in $\varepsilon$ as the other terms (i.e. the order zero, although it is really a second-approximation formula: the first approximation gives simply $du^i/dT = 0$). In contrast, in the pN equation of motion for a mass point, Eq. (72), $U_{,i}$ represents the first approximation to the acceleration, of order zero in $\varepsilon$, and the other, pN terms, are of the order $\varepsilon^2$.

**iii)** *Transition to a moving frame. Application to the effects of a weak gravitational field on light rays*
In order that the pN motion may be considered as a perturbation of the problem in classical celestial mechanics, one has to work in the mass-center frame, as in classical mechanics. Let $\mathbf{V}(T)$ be the current absolute velocity of the mass-center. We define the mass-center frame $\mathsf{E_V}$ as the frame that undergoes a pure translation, with velocity $\mathbf{V}$, with respect to the ether frame $\mathsf{E}$ ($\mathbf{V}$ may vary with $T$, although very slowly in the case of the solar system). We may pass from $\mathsf{E}$ to $\mathsf{E_V}$ by a Lorentz transformation of the flat space-time metric $\boldsymbol{\gamma}^0$ (that one whose line element is $(ds^0)^2 = c^2\, dT^2 - dx^i dx^i$ in Cartesian coordinates for the Euclidean space metric $\mathbf{g}^0$). In a such transformation, the components of the velocity $\mathbf{u} = d\mathbf{x}/dT$ and the acceleration $\mathbf{a} = d\mathbf{u}/dT$ transform by the classical formulas of *special* relativity (as far as we assume, as it is very reasonable, that the acceleration $d\mathbf{V}/dT$ plays no role at the pN approximation). Expanding the corresponding transformations of Eqs. (71) and (72) gives the pN equations of motion in the uniformly moving frame $\mathsf{E_V}$, for a photon and for a mass point.

However, at the pN approximation, a photon follows a null geodesic of the physical space-time metric $\boldsymbol{\gamma}$, and, by the Lorentz transformation, the components of this space-time tensor transform thus like a (twice covariant) tensor, of course: this gives an alternative way to get the pN equations of motion for a photon in the frame $\mathsf{E_V}$. Neglecting $O(1/c^4)$ terms, and except for a $O(1/c^3)$ term in $\gamma'_{0i}$ (which does not play any role in the pN expansion of the null geodesic equation), the new components $\gamma'_{\mu\nu}$ (in the frame $\mathsf{E_V}$) depend only on the "Newtonian" potential $U$, by just the same equations (65) and (66) as they do in the preferred frame. Now the effects of gravitation on light rays are always calculated at the pN approximation and using the additional assumption of a *spherical* and *static* gravitational field [18, 26, 38, 40]. In the case of spherical symmetry, the "Newtonian" potential is simply $U = GM^0/r$ with $M^0 \equiv \int \rho^0\, dV$: then, Eqs. (65) and (66) represent just the pN expansion of Schwarzschild's exterior metric. Whereas this spherical potential is, in general, not constant in the frame $\mathsf{E}$, it is indeed constant in the frame $\mathsf{E_V}$ which moves with the spherical massive body (e.g. the Sun) that creates the relevant field, so that the geodesic equation is really that deduced from Schwarzschild's metric. We conclude that, *to just the same level of approximation as in the pN approximation of GR,* in particular neglecting the $\gamma_{0i}$ components of the metric, which are $O(1/c^3)$, *the present theory predicts exactly the same gravitational effects on light rays as does standard general relativity, and this is true accounting for the preferred-frame effects (which do not appear at this approximation).* (The arguments are detailed in Ref. 6 .)

**iv)** *Remarks on the adjustment of astrodynamical constants and the preferred-frame effects*
In contrast to the pN acceleration (71) for a photon, which is invariant by a Lorentz transformation of the flat metric (provided the velocity $V$ of the moving frame is compatible with the pN approximation, i.e. such that $V/c = O(\varepsilon)$), the pN acceleration for a mass point, Eq. (72), is *not* invariant. For a mass point, the pN acceleration in the moving frame is, in space vector form,

$$d\mathbf{u}/dT = (d\mathbf{u}/dT)_0 + (d\mathbf{u}/dT)_\mathbf{V}, \qquad (73)$$

where $(d\mathbf{u}/dT)_0$ is given, in the Lorentz-transformed Galilean coordinates $x'^\mu$ for the flat metric, by just the same formula (72), but with the indices and derivatives referring to the $x'^\mu$ coordinates, and where $(d\mathbf{u}/dT)_\mathbf{V}$ is a sum of a few terms, each term containing explicitly the velocity $\mathbf{V}$. Thus, the theory does predict preferred-frame effects for mass particles such as the planets. It is not difficult (though rather tedious) to calculate the magnitude of the effect, on Mercury's perihelion motion, of the $(d\mathbf{u}/dT)_\mathbf{V}$ part: for an absolute velocity of the order $V \approx 10^{-3}\,c$, it is comparable to that of the "relativistic" correction based on the motion in the Schwarzschild field. On the other hand, for $V = 0$, and for a spherical body, the $(d\mathbf{u}/dT)_0$ acceleration is just that derived from Schwarzschild's metric at the pN level, and which gives the "miraculous" 43" per century. Since $V \approx 10^{-3}\,c$ is much more plausible than $V \approx 0$, the theory would appear, at first sight, unable to explain Mercury's residual advance in perihelion. A number of other alternative theories are in the same situation, and it is often concluded on this basis that such theories are to be rejected [40].

However, one may consider that things are less simple than this. The main point is that, in classical celestial mechanics, the astrodynamical constants such as the Newtonian masses $M_i^{\,N}$ of the celestial bodies (or rather the products $GM_i^{\,N}$) are not *measured* (one cannot weigh a planet!), instead they are *adjusted to best fit the observations*. As an illustrative example, if a system of two celestial bodies with masses $M_1^{\,N}$ and $M_2^{\,N}$ is considered as isolated (the Sun and Venus, say), Kepler's third law (which is exact *in the frame of NG* for the case of a two-bodies problem) allows deduction of $G.(M_1^{\,N} + M_2^{\,N})$ from the observed period $T$ and semi-major axis $a$ of the relative motion:

$$G.\left(M_1^{\,N} + M_2^{\,N}\right) = \left(\frac{2\pi}{T}\right)^2 a^3. \qquad (74)$$

In reality, no couple of celestial bodies is exactly isolated, and perturbation theory allows to modify the astrodynamical constants by successive corrections, always remaining in the frame of Newtonian theory (NG). With modern computers, one may calculate hundreds of Newtonian parameters by a global fitting based on a numerical approximation scheme of NG and using hundreds of input data from various observation devices (*cf.* Müller *et al.* [28]).

We know that the correct theory of gravitation is not NG. Let us assume for a moment that it is instead some non-linear theory (T), the equations of which reduce to NG in the first approximation. The pN approximation of theory (T) will introduce, in addition to the first-approximation masses $M_i^{\,0}$ (and higher-order multipole moments of the mass density $\rho^{\,0}$ of the first approximation, etc.), some pN parameters such as the pN corrections to the active masses, say $M_i^{\,1}$. Obviously, the fitting of observational data must now be carried out *within the pN approximation of theory (T)*. There is simply no reason that the first-approximation masses $M_i{}^0$, which are obtained (together with the corrections $M_i{}^1$) by this fitting, coincide with the "effective Newtonian" masses $M_i{}^N$, that are obtained by fitting the observational data *within NG*. In other words, Newtonian astrodynamics *in praxi* does not coincide with the first approximation of the non-linear theory (T) − unless the pN approximation of theory (T) makes negligible corrections to NG. Specifically, if the theory predicts significant preferred-frame effects, and if turns out that these effects are really present in Nature, the masses $M_i{}^N$ will be affected by the preferred-frame effects (in particular, they would be found different for two otherwise identical systems moving at different absolute velocities), whereas the first-order masses $M_i{}^0$ will not be affected − at least under the assumption that the pN approximation of theory (T) describes the system up to a negligible error.

To investigate the consequence of this state of things, let us call $D_j{}^0$ and $D_j{}^1$ the first-order prediction and the pN correction for an observational data $D_j$ such as Jupiter's period or Mercury's advance in perihelion: thus, $D_j{}^0$ is obtained by using the "true" values of the first-approximation parameters, e.g. the first-order masses $M_i{}^0$. (The values $M_i{}^0$, $M_i{}^1$, and so on, obtained by a least-squares fitting of some subset of the parameters $D_j$ using the the pN approximation of theory (T), are indeed the "true" ones if this pN approximation describes the system up to a negligible error.) Since Newtonian astrodynamics *in praxi* tries to accommodate a subset of the observational data $D_j$, which are affected by pN effects $D_j{}^1$, by using merely the first-approximation formulas, one may expect that, in general, it will lead to values for the first-order parameters and predictions, $M_i{}^N$ and $D_j{}^N$, differing from the true first-order values $M_i{}^0$ and $D_j{}^0$ by quantities whose order of magnitude is that of the pN corrections $M_j{}^1$ and $D_j{}^1$. But, in turn, the pN calculation based on the wrong values of the first-order parameters, such as the wrong first-order masses, thus $M_i{}^N$ instead of $M_i{}^0$, will predict pN corrections $D'_j{}^1$ differing from the true ones $D_j{}^1$ by *third-order* quantities, which are likely to be negligible. As a consequence, this wrong pN calculation, using the effective Newtonian masses $M_i{}^N$ instead of the true first-order masses $M_i{}^0$, will give pN predictions $D'_j{}^{pN}$:

$$D'_j{}^{pN} \equiv D_j{}^N + D'_j{}^1 \approx D_j{}^N + D_j{}^1 \neq D_j{}^0 + D_j{}^1 \equiv D_j{}^{pN}, \qquad (75)$$

differing from the correct pN predictions $D_j{}^{pN}$ by quantities of the same order of magnitude as the pN corrections. In summary, if the *correct* pN approximation of theory (T) turns out to describe celestial mechanics in the solar system up to a negligible error, then the *incorrect* pN procedure, that consists in assigning to the first-order parameters the values of the Newtonian effective parameters, is likely to give rather poor predictions, which might well be less accurate than Newtonian theory, i.e. less accurate than the first-approximation alone.

The foregoing argument applies *a priori* also to *general relativity* (although there is no preferred-frame effect in GR, there are of course non-Newtonian, "relativistic" effects). As a relevant example, the standard pN approximation of GR, using the harmonic gauge condition [38], introduces, besides the "Newtonian" (first-approximation) potential $U$ (denoted $-\phi$ by Weinberg [38]), still another scalar potential obeying a Poisson equation. Namely, it introduces the pN potential denoted $\psi$ by Weinberg [38], which plays exactly the same role as $A/2$ in the present theory (cf. Eqs. (66) and (69) hereabove). Weinberg writes explicitly that "we can take account of $\psi$ by simply replacing $\phi$ everywhere by $\psi + \phi$ " [or $\psi/c^2 + \phi$, since Weinberg sets $c = 1$], consistently with our statement that Newtonian astrodynamics *in praxi* does not coincide with the first approximation of the non-linear theory, here GR.

Coming back to the proposed theory, the foregoing argument is not a proof, of course, that the theory does correctly explain all minute discrepancies between classical celestial mechanics and observations in the solar system. This argument merely suggests that the present theory should not be *a priori* rejected on the basis that it predicts preferred-frame effects in celestial mechanics – the more so as this theory correctly explains the gravitational effects on light rays, which are the most striking and the best established predictions of GR.

## 9. Conclusion

A rather complete theory of gravitation may be built from an extremely simple heuristics, already imagined by Euler, and that sees gravity as Archimedes' thrust in a perfectly fluid ether. The theory thus obtained is very simple also, as it is a scalar bimetric theory in which the metric effects of gravitation are essentially the same as the metric effects of uniform motion, and with the field equation being a modification of d'Alembert's equation. Perhaps the most important finding is that Newton's second law may be extended to a general space-time curved by gravitation, in a way that is both consistent and seemingly compelling. This extension is not restricted to point particles: it applies also to any kind of continuous medium, in fact it is Newton's second law for the electromagnetic *field continuum* that gives the Maxwell equations of the theory.

Although the general extension of Newton's second law is unique, its exact expression still depends on which assumption is stated for the gravity acceleration. It is at this point that the ether heuristics makes its demarcation from the logic of space-time which leads to Einstein's geodesic assumption and Einstein's equivalence principle. The ether heuristics leads indeed to postulate that the gravity acceleration depends only on the local state of the imagined fluid, whereas geodesic motion can be true in a general, time-dependent metric, only if the gravity acceleration depends also on the velocity of the particle – and this, in such a way that the velocity-dependent part of the gravitation force does work [5]. Hence, the here-assumed gravity acceleration is in general incompatible with geodesic motion and, in connection with this, it leads to an equation for continuum dynamics, Eq. (37), which differs from the equation that goes with geodesic assumption and Einstein's equivalence principle, Eq. (24). It also enforces that the theory is a preferred-frame theory, which is independently enforced by the scalar character of the theory. It is interesting to note that the non-Newtonian effects, as well as the preferred-frame character, come once a non-zero compressibility is attributed to the ether, just like in Dmitriev's theory [16].

The alternative continuum dynamics implies that, in a variable gravitational field at a high pressure, matter is produced or destroyed, by a reversible exchange with the gravitational field, Eq. (44). The tenuous amounts seem compatible with the experimental evidence on "mass conservation". An experimental confirmation or refutation of this new form of energy exchange would be extremely interesting. In the same way, the alternative continuum dynamics leads to Maxwell equations that contain the possibility of electric charge production/destruction in an electromagnetic field *subjected to* a variable gravitational field. However, the amounts cannot be estimated without having recourse to a numerical work, which is yet to be done. This is obviously a dangerous point for the theory, although it is also an interesting one.

As to the classical tests, it has been shown that no preferred-frame effect exists for light rays at the first post-Newtonian approximation, and that in fact the gravitational effects on light rays are correctly predicted. It has also been shown that preferred-frame effects do exist in celestial mechanics, and it has been argued that this does not kill the theory. Furthermore, such effects might play an interesting role at larger scales, e.g. they might contribute to explain rotation curves in galaxies. We mention also that, contrary to general relativity, the theory predicts a *bounce* instead of a *singularity* for the gravitational collapse of a dust sphere [3]; and that, like general relativity, it leads to a "quadrupole formula" that rules the energy *loss*, thus the correct sign, for a gravitating system with rapidly varying field (this is less shortly outlined in Ref. 3).

Finally, it has not been attempted to attack the impressive task of linking the present theory with quantum mechanics and quantum field theory. Only some preliminary remarks may be done: first, the abandon of the relativity principle (in the presence of a gravitational field) and its replacement by a preferred frame (presumably the average rest frame of the universe/ether) change drastically the problematic. In particular, it makes sense, in this framework, to assign a non-punctual (albeit fuzzy) extension to particles. Further, the quantum non-separability, and the unity of matter and fields, are strongly compatible with the heuristic interpretation of particles as organized flows in the fluid ether. Another remark is that quantum theory is based on the Hamiltonian/Lagrangian formalism, and that the present theory is not amenable to this formalism, except for a constant gravitational field. But Newton's second law is a more general tool than these variational principles, since it makes sense also in non-conservative situations. However, the existence, in the present theory, of a local conserved energy (made of the gravitational energy, plus the energy of matter and non-gravitational fields), should play an important role.

## Acknowledgement

I am grateful to Prof. E. Soós (Institute of Mathematics of the Romanian Academy, Bucharest) and to Prof. P. Guélin (Laboratoire "Sols, Solides, Structures", Grenoble) for helpful discussions.

## References

1. M. Arminjon, *A theory of gravity as a pressure force. I. Newtonian space and time*, Rev. Roum. Sci. Tech.- Méc. Appl., **38** (1993), 3-24.
2. M. Arminjon, *A theory of gravity as a pressure force. II. Lorentz contraction and 'relativistic' effects*, Rev. Roum. Sci. Tech.- Méc. Appl., **38** (1993), 107-128.
3. M. Arminjon, *Scalar theory of gravity as a pressure force*, Rev. Roum. Sci. Tech.- Méc. Appl., **42** (to appear in 1997).
4. M. Arminjon, *Energy and equations of motion in a tentative theory of gravity with a privileged reference frame*, Arch. Mech., **48** (1996), 25-52.

5. M. ARMINJON, *On the extension of Newton's second law to theories of gravitation in curved space-time*, Arch. Mech., **48** (1996), 551-576.
6. M. ARMINJON, *Post-Newtonian approximation of a scalar theory of gravitation and application to light rays*, Rev. Roum. Sci. Tech.- Méc. Appl. (to appear).
7. M. ARMINJON, *Continuum dynamics and mass conservation in a scalar theory of gravitation* (submitted).
8. M. ARMINJON, *Continuum dynamics and the electromagnetic field in a preferred-frame theory of gravitation* (submitted).
9. L. BRILLOUIN, *Les tenseurs en mécanique et en élasticité*, Masson, Paris (1949).
10. G. BUILDER, *Ether and relativity*, Austr. J. Phys., **11** (1958), 279-297.
11. G. BUILDER, *The constancy of the velocity of light*, Austr. J. Phys., **11** (1958), 457-480.
12. M.O. CALVÃO, J.A.S. LIMA and I. WAGA, *On the thermodynamics of matter creation in cosmology*, Phys. Lett. A, **162** (1992), 223-226.
13. C. CATTANEO, *General relativity: relative standard mass, momentum, energy and gravitational field in a general system of reference*, Nuovo Cimento, **10** (1958), 318-337.
14. S. CHANDRASEKHAR, *The post-Newtonian equations of hydrodynamics in general relativity*, Astrophys. J., **142** (1965), 1488-1512.
15. S. CHANDRASEKHAR, *Conservation laws in general relativity and in the post-Newtonian approximations*, Astrophys. J., **158** (1969), 45-54.
16. V.P. DMITRIEV, *Elastic model of physical vacuum*, Mech. Solids (New York), **26** (1992), 60-71. [Transl. from Izv. Akad. Nauk SSSR, Mekhanika Tverdogo Tela, **26** (1992), 66-78.]
17. M.C. DUFFY, *Ether, cosmology and general relativity*, Proc. 5$^{th}$ Conf. "Physical Interpretations of Relativity Theory", Supplementary papers, British Society for the Philosophy of Science (to appear in 1997).
18. V. FOCK, *The theory of space, time and gravitation* (2$^{nd}$ English edition), Pergamon, Oxford (1964).
19. F. HOYLE and J.V. NARLIKAR, *Mach's principle and the creation of matter*, Proc. Roy. Soc., **A 273** (1963), 1-11.
20. L. JÁNOSSY, *The Lorentz principle*, Acta Physica Polonica, **27** (1965), 61-87.
21. L. JÁNOSSY, *Theory of relativity based on physical reality*, Akademiai Kiado, Budapest (1971).
22. R.T. JANTZEN, P. CARINI and D. BINI, *The many faces of gravitoelectromagnetism*, Ann. Physics (New York), **215** (1992), 1-50.
23. L. LANDAU and E. LIFCHITZ, *Théorie des champs* (3$^{rd}$ French edition), Mir, Moscow (1970).
24. A. LICHNEROWICZ, *Eléments de calcul tensoriel*, Armand Colin, Paris (1950).
25. P. MAZILU, *Missing link in the theory of gravity interaction*, Proc. 6$^{th}$ Marcel Grossmann Meet. Gen. Relat. (H. Sato & T. Nakamura, eds.), World Scientific, Singapore (1992).
26. C.W. MISNER, K.S. THORNE and J.A. WHEELER, *Gravitation*, Freeman, San Francisco (1973).
27. C. MØLLER, *The theory of relativity*, Clarendon Press, Oxford (1952).
28. J. MÜLLER, M. SCHNEIDER, M SOFFEL and H. RUDER, *New results for relativistic parameters from the analysis of Lunar Laser Ranging measurements*, in Relativistic Gravity Research (J. Ehlers and G. Schäfer, eds.), Springer, Berlin-Heidelberg-New York (1992), pp. 87-99.
29. M.F. PODLAHA, *Gravitation and the theory of physical vacuum II*, Ind. J. Theor. Phys., **32** (1984), 103-112.
30. M.F. PODLAHA and T. SJÖDIN, *On universal fields and de Broglie's waves*, Nuov. Cim., **79B** (1984), 85-92.
31. H. POINCARÉ, *Sur la dynamique de l'électron*, Rendic. Circ. Matemat. Palermo, **21** (1906), 129-176.

32. I. PRIGOGINE, J. GEHENIAU, E. GUNZIG and P. NARDONE, *Thermodynamics and cosmology*, Gen. Relat. Gravit., **21** (1989), 767-776.
33. S.J. PROKHOVNIK, *The logic of special relativity*, Cambridge University Press, Cambridge (1967).
34. S.J. PROKHOVNIK, *Light in Einstein's universe*, Reidel, Dordrecht (1985).
35. L. ROMANI, *Théorie générale de l'univers physique* (2 vols.), Librairie Scientifique A. Blanchard, Paris (Vol. 1: 1975, Vol. 2: 1976).
36. T. SJÖDIN, *A scalar theory of gravitation: motion in a static spherically symmetric field*, Found. Phys. Lett., **3** (1990), 543-556.
37. J. L. SYNGE, *Relativity, groups and topology* (B. De Witt and C. De Witt, eds.), Gordon and Breach, New York - London (1964), pp. 79-88.
38. S. WEINBERG, *Gravitation and cosmology*, J. Wiley & Sons, New York (1972).
39. E.T. WHITTAKER, *A history of the theories of aether and electricity*, Vol. I, Th. Nelson & Sons, London (1953).
40. C.M. WILL, *Theory and experiment in gravitational physics* ($2^{nd}$ edn.), Cambridge University Press, Cambridge (1993).
41. Y.Z. ZHANG, *Test theories of special relativity*, Gen. Rel. Grav., **27** (1995), 475-493.